\documentclass[review]{elsarticle}
\usepackage[margin=1.5in]{geometry}
\usepackage{amsthm,amssymb,amsmath}

\usepackage[english]{babel}
\usepackage[utf8]{inputenc}
\usepackage{amsmath}
\usepackage[dvipsnames]{xcolor}
\usepackage{graphicx}
\usepackage[colorinlistoftodos]{todonotes}

\usepackage{placeins}

\usepackage{booktabs}

\usepackage{subcaption}
\usepackage{setspace}

\usepackage[hyphens]{url}
\usepackage{hyperref}
\hypersetup{hidelinks=ture}
\usepackage[hyphenbreaks]{breakurl}

\usepackage{multirow}
\usepackage{float}
\usepackage{comment}

\usepackage{xspace}
\newcommand\ie{i.\,e.\xspace}
\newcommand\eg{e.\,g.\xspace}

\newcommand\US{U.\,S.\xspace}

\usepackage[left]{lineno}

\makeatletter
\def\ps@pprintTitle{%
     \let\@oddhead\@empty
     \let\@evenhead\@empty
     \def\@oddfoot{\footnotesize\itshape}%
     \let\@evenfoot\@oddfoot
     }
\makeatother

\begin{document}

\begin{frontmatter}

\title{New threats to society from free-speech social media platforms}

\author[lmu,mcml]{Dominik Bär}
\author[jlu]{Nicolas Pröllochs}
\author[lmu,mcml]{Stefan Feuerriegel\corref{cor1}}
\cortext[cor1]{Corresponding author, email address: \url{feuerriegel@lmu.de}}
\address[lmu]{LMU Munich, Germany}
\address[mcml]{Munich Center for Machine Learning, Germany}
\address[jlu]{JLU Giessen, Germany}

\begin{abstract}
In recent years, several free-speech social media platforms (so-called ``alt-techs'') have emerged, such as Parler, Gab, and Telegram. These platforms market themselves as alternatives to mainstream social media and proclaim ``free-speech'' due to the absence of content moderation, which has been attracting a large base of partisan users, extremists, and supporters of conspiracy theories. In this comment, we discuss some of the threats that emerge from such social media platforms and call for more policy efforts directed at understanding and countering the risks for society. 
\end{abstract}
\end{frontmatter}

\clearpage

\section{Introduction}


The attack on the \US Capitol on January 6, 2021, led to the killing of 5 people, turning it into an unprecedented attack on democracy \citep{Otala.2021}. Recent evidence documented that the attack was partially coordinated through {free-speech social media platforms} (so-called ``alt-techs'') \citep{Otala.2021}. Alt-techs such as Parler, Gab, and Telegram constitute new information, communication, and socialization ecosystems, which lack content moderation, and thus provide a parallel online space for ideas that are outside the boundaries of speech permitted on mainstream platforms. As a result, alt-techs facilitate the circulation of biased, inaccurate, misleading, and conspiratorial content at unprecedented levels and, furthermore, have attracted large numbers of partisan users and extremists. As seen in the example of the attack on the \US Capitol and evidence from other violent incidents, the recent emergence of alt-techs might even have dangerous effects on the offline world, including democracies as a whole.

\section{What are alt-techs?} 


Alt-techs have emerged as a new social media phenomenon \citep{McIlroyYoung.2019}. Prominent examples include Parler (\url{www.parler.com}), Gab (\url{www.gab.com}), 4chan (\url{www.4chan.org}), 8chan (\url{www.8ch.net}), Voat (\url{www.voat.co}), Gettr (\url{www.gettr.com}), BitChute (\url{www.bitchute.com}), Telegram (\url{www.telegram.org}), Discord (\url{www.discord.com}), and Mastodon (\url{www.joinmastodon.org}). These are part of the broader {alt-tech} ecosystem which provides alternative websites, platforms, and services specifically targeted toward certain, often partisan or fringe communities. The features of alt-techs are typically not innovative in themselves but rather offer services largely similar in functionality to those found on mainstream platforms (\eg, Twitter, Facebook). However, different from mainstream social media, alt-techs self-proclaim as ``free-speech'' platforms and, because of that, eschew content moderation so that users can post content that would not be permitted on mainstream social media.

The reasons for the emergence of alt-techs are many-faceted. In recent years, partisan communities have developed a narrative claiming that speech on mainstream platforms is being ``censored'' for failure to be ``politically correct'' \citep{Papasavva.2021}. Such views can be partially attributed to the increasing content moderation efforts from mainstream platforms to limit misinformation, conspiracy theories, and hate speech. For instance, when Twitter and other platforms increased efforts to ban malicious accounts or flag misinformation, many conservative users migrated to Gab and Parler \citep{Ali.2021, Otala.2021}. Conservative thought leaders also jumped on the bandwagon by endorsing, in particular, Parler as an alternative to mainstream social media \citep{Aliapoulios.2021}. Similarly, many liberals recently endorsed Mastodon as a Twitter alternative being dissatisfied with Elon Musk's new content moderation policy \cite{Lima.2022}. As such, migrating to alt-techs can be seen as a political statement. Overall, a mixture of content restrictions, deplatforming, and bans imposed by ``big tech'' has contributed to the popularity of alt-techs.


A particular characteristic of alt-techs is their user base. Especially during the 2020 \US presidential election, many alt-techs have witnessed stark growth; \eg, Parler counted around one million users in June 2020 but more than 13 million only six months later in January 2021 \citep{Aliapoulios.2021}. The user base of alt-techs tends to be rather homogeneous: a large share of users even self-identify as partisans, extremists, or conspiracy theorists \citep{Aliapoulios.2021, Zannettou.2018}. Mirroring the user base, the content on alt-techs covers---to a large extent---extreme viewpoints or conspiracy theories. For example, Gab features high levels of hate speech, toxic, and antisemitic content  \citep{Zannettou.2018, McIlroyYoung.2019, Ali.2021}. Furthermore, on Parler, hashtags such as \#qanon are widespread and posts frequently link to websites known for spreading misinformation (\eg, \url{www.breitbart.com}) \citep{Aliapoulios.2021}. As such, alt-techs form ideologically-driven ecosystems where opposing viewpoints are largely absent \citep{Freelon.2020}. This can be problematic: Even in mainstream social media, where users are regularly exposed to diverse ideological content, echo chambers of users sharing similar worldviews have led to increased political polarization \citep{Finkel.2020}. In the absence of opposing views, alt-techs are likely to accelerate the growth of echo chambers and further benefit polarization and radicalization \citep{Freelon.2020}.

\FloatBarrier
\section{Why alt-techs are a threat to society}

Alt-techs pose direct concerns for society, particularly because they have been frequently associated with violent incidents in the offline world. For instance, the offender in the Buffalo shooting on May 14, 2022, allegedly shot 10 people in a racist attack inspired by content on 4chan \cite{Collins.2022}. Furthermore, Gab was the main communication channel during the Pittsburgh synagogue shooting on October 27, 2018, a white supremacist terrorist attack that caused 11 people to lose their lives \citep{McIlroyYoung.2019}. A similar role for alt-techs has been observed for the attack on the \US Capitol on January 6, 2021. Rioters communicated via Parler on how to evade police forces or smuggle weapons into the capitol \cite{Otala.2021}. As such, Parler is nowadays regarded as the primary communication channel of the rioters before and during the Capitol attack \citep{Otala.2021, Jakubik.2023}. 


Alt-techs further provide fertile grounds for misinformation and even disinformation. This includes deliberately deceptive content, with which users seek to spread conspiracy theories, false rumors, hoaxes, and inflammatory opinions to promote their own ideological viewpoints, decrease trust in mainstream institutions, and win others as followers \citep{Freelon.2020}. The proliferation of such content on alt-techs may be partially attributed to their promise of ``uncensored'' speech. Specifically, the lack of content moderation allows spreaders of misinformation to fill the curiosity gap by sensationalizing ``censored'' content (\ie, ``what they don't want you to know''). Another driver that may promote the spread of misinformation is the homogeneity of users. Due to the segregation, confrontation and exchange with other views are rare and thus could remove skepticism that spreaders of misinformation may face on mainstream social media. From a societal perspective, the proliferation of misinformation on alt-techs is alarming as it may undermine the concepts of truth and reality among users, which might directly affect the offline world. For instance, the alleged shooter of the Buffalo shooting was suspected to be motivated by the ``replacement theory,'' a conspiracy theory claiming that a cabal attempts to replace white Americans with non-white people that circulated on 4chan \cite{Collins.2022}.

\section{What we know about alt-techs} 


Despite imminent concerns, research focusing on a better understanding of alt-techs has remained scarce. 
One literature stream seeks to generate a better understanding of \textbf{\emph{who}} uses alt-techs. For example, many users migrated to Parler from Twitter during the 2020 \US presidential election \citep{Otala.2021}, and, on top of that, the average user on Parler has strong partisan views \citep{Aliapoulios.2021}. Similarly, many users who previously violated Twitter's community guidelines and then migrated to Gab contributed to a more toxic and radical ecosystem \citep{Ali.2021}. Besides that, users of alt-techs commonly advocate conspiracy theories \citep{Aliapoulios.2021, Zannettou.2018}. For example, prior research has shown that users frequently discuss QAnon on Parler, 4chan, and Voat \citep{Aliapoulios.2021, Aliapoulios.2022}.

Another literature stream studies \textbf{\emph{what}} is shared on alt-techs and, for that purpose, compares alt-techs against mainstream social media. For example, content shared on Parler is significantly different from that shared on Twitter, especially before and after the attack on the \US Capitol: many Parler users expressed a less negative sentiment and lower levels of guilt compared to users on Twitter \citep{Jakubik.2023}. Furthermore, content on alt-techs such as 4chan, BitChute, and Gab contains more hate speech than mainstream social media \citep{Freelon.2020}. For example, antisemitic hate speech was widely shared on Gab in response to the deadly Pittsburgh synagogue shooting \citep{McIlroyYoung.2019}. Moreover, websites known for spreading misinformation and partisan content, such as  \url{www.thegatewaypundit.com} or \url{www.breitbart.com}, are among the most frequently shared links on Parler and Gab \citep{Aliapoulios.2021, Zannettou.2018}. Overall, this presents increased risks that make users of alt-techs vulnerable to polarization and radicalization compared to mainstream social media.

\section{The way forward}


There is an urge for policy-relevant research to better understand alt-techs as newly emerging phenomena. The knowledge available to policymakers often builds upon anecdotal evidence rather than empirical evidence, which may limit assessments that are rigorous, representative, and comprehensive. Due to differences in the user base and content dynamics, earlier findings from mainstream social media might no longer apply.


Interdisciplinary research combining both computational and social science is important to unravel the underlying mechanisms of alt-techs and inform policies to counter emerging threats to society. On the one hand, computational models help social scientists to develop theories that describe the mechanism of information diffusion in social networks. On the other hand, theories from social science help computer scientists by informing the design of computational methods, so that these are effective in capturing the underlying data-generating process (\eg, to account for potential confounding, avoid bias, and develop algorithms that are perceived as trustworthy and fair). Theories from social science can further explain why and where countermeasures such as de-bunking are effective (\eg, using theoretical concepts such as social norms, in/out-group members, etc.). Finally, an interdisciplinary research agenda is essential to establish ethical guidelines for scientists studying sensitive personal information or even individual users of alt-techs.

To address risks emerging alt-techs, we advocate impactful research along three dimensions, \ie, users, content, and society (see Table~\ref{tab:examples_rq}). At the user level, more effort is needed to characterize the different groups and identify behavioral factors that promote migration tendencies. At the content level, advanced computational methods, \eg, from natural language processing, combined with cross-platform analyses, can provide tools to systematically analyze the proliferation of hate speech, conspiracy theories, and misinformation on alt-techs, as well as their role during specific offline events. At the societal level, there is a demand to evaluate the effect of alt-techs on the segregation, polarization, and radicalization in the offline world, as well as voter behavior and disinformation in war and conflict (\eg, coordinated Russian propaganda in the Russia-Ukraine war).


There is further a need for evidence-based policies that mitigate the risks alt-techs pose to the functioning of modern societies. However, alt-techs have only little incentive to adopt stricter content moderation policies as the lack of such policies is the main reason for their popularity. Here, research such as that discussed above can inform new regulations that enforce alt-techs to establish a level of content moderation that complies with democratic law (rather than putting the power into the hand of private companies). For instance, the Network Enforcement Act (NetzDG) in Germany and the Cybersecurity Law in China force platforms to counter misinformation and hate speech by imposing severe penalties for misconduct. In addition, public pressure (e.g., from non-governmental organizations and the media) can have a critical impact on the platforms’ businesses. As an example, Parler’s alleged role in the storming of the U.S. Capitol in January 2021 led to the removal of the Parler app from both the Google Play Store and Apple’s App Store, as well as Amazon stopping to host the website, which seized Parler from operating for multiple months.


Regulatory policies and public pressure may help in incentivizing platforms to foster academic research. For example, in light of the EU Digital Services Act and pressure from \US politics, Twitter and Meta launched programs such as the Twitter Transparency Center or the Meta Ad Library to enhance transparency and accessibility of internal data for researchers. We expect that similar efforts will also be effective for alt-techs (\eg, Germany's NetzDG compelled Telegram to cooperate with law enforcement agencies since, otherwise, the platform risks being banned from app stores). Along these lines, policymakers can enforce transparency to enable enhanced data access or even mandate platform providers to implement measures for content moderation (\eg, counter-speech, fact-checking) and have their effectiveness evaluated through third-party bodies (\eg, by researchers conducting large-scale field experiments). Overall, this might contribute to a less radical ecosystem and mitigate the safety risk originating from alt-techs.     


To sum up, alt-techs have emerged from niche existence to social media ecosystems that host a large base of partisan users, extremists, and conspiracy theorists. A better understanding of alt-techs is critical as user counts reach new highs and new platforms such as Truth Social (\url{www.truthsocial.com}) by Donald Trump emerge. Hence, more effort is needed to inform evidence-based policies and other regulations that could mitigate the risks alt-techs pose to society. 

\begin{table}[h!]
    \centering
    \footnotesize
    \onehalfspacing
    \begin{tabular}{p{0.15\textwidth}p{0.8\textwidth}}
        \toprule
         \textbf{Area} & \textbf{Research questions (examples)} \\
         \midrule
         \multirow{4}{*}{\vspace{-1.3em} User} & \emph{What are characteristics of users and their online behavior on alt-techs?} \\
         \cmidrule(lr){2-2}
         & \emph{What discriminates different user groups on alt-techs (\eg, partisans, extremists, conspiracy theorists)?}\\
         \cmidrule(lr){2-2}
         & \emph{Which societal biases exist on alt-techs and how are they reinforced on the platforms?} \\
         \midrule
         \multirow{4}{*}{\vspace{-1em} Content} & \emph{What content is posted to what extent on alt-techs (\eg, topics, media, links)?} \\
         \cmidrule(lr){2-2}
         & \emph{What explains the proliferation of misinformation?} \\
         \cmidrule(lr){2-2}
         & \emph{How effective are techniques for content moderation such as counter-speech or misinformation flags on alt-techs?} \\
         \midrule
         \multirow{4}{*}{\vspace{-1.5em} Society} & \emph{To what extent are segregation, polarization, and radicalization present on alt-techs?} \\
         \cmidrule(lr){2-2}
         & \emph{What is the impact of alt-techs on election outcomes?}\\
         \cmidrule(lr){2-2}
         & \emph{What are societal risks of alt-techs and what are effective countermeasures?}\\
         \bottomrule
    \end{tabular}
    \caption{Key research questions to better understand alt-techs.}
    \label{tab:examples_rq}
\end{table}

\newpage

\bibliographystyle{naturemag}
\bibliography{literature}

\end{document}